# Shrinking-shifting and Amplifying-shifting Device Using Transformation Optics

Hamza Ahmad Madni,[1,2,5] Muhammad Musavir Bilal,[3] Farrukh Jaleel,[4] Ahmed Sohaib,[1] and Wei Xiang Jiang[2,6]

[1] *Department of Computer Engineering, Khwaja Fareed University of Engineering and Information Technology, Rahim Yar Khan 64200, Pakistan.*
[2] *State Key Laboratory of Millimeter Waves, Department of Radio Engineering, Southeast University, Nanjing 210096, China*
[3] *College of Information Science and Engineering, Yanshan University, Qinhuangdao, Hebei 0066004 China.*
[4] *Department of Chemistry, Khwaja Fareed University of Engineering and Information Technology, Rahim Yar Khan 64200, Pakistan.*
[5] *Corresponding author: 101101770@seu.edu.cn*
[6] *Corresponding author: wxjiang81@seu.edu.cn*

**Based on transformation optics (TO), this paper uses geometric divisions and linear coordinate transformations to design "shrinking-shifting - and reshaping", and "amplifying-shifting - and reshaping" devices. The proposed devices can reshape the sizes and locations of the wrapped-objects inside the core-region. The shrinking-shifting device shrinks the larger object into a smaller one and shifts it to different location, whereas the shrinking-reshaping device can generate a smaller-size image with different shape located at different location. In contrast to previously designed shrinking devices, the real object wrapped inside the proposed core-region and the transformed object contains the same material properties, and the location-shifting is another feature. Here, the shifting-region is located inside the physical-space boundaries to achieve the non-negative, homogeneous, and anisotropic material properties of the proposed device, which are easier for real implementations. Thus, we further verified this concept with the amplifying-shifting and -reshaping devices for visually transformation of smaller object into bigger one placed at different location and position. We also applied active scatterer to further validate the working functionality of proposed devices. In addition, the proposed devices behave like the concentrator and (or) rotator effect in the absence of any scatterer. Our findings highlight the role of TO, suggesting directions for future research on bi-functional devices that will be useful for shrinking and amplifying devices, illusion optics, camouflage, and object protection etc.**

## 1. INTRODUCTION

Transformation Optics (TO) is considered as a powerful tool to establish the equivalence between constitutive parameter distribution and spatial transformation in order to control the electromagnetic (EM) field distribution [1–4]. In this regard, the mapping technique of transformation from virtual-space to physical-space using coordinate-transformation recipe is very important in achieving the constitutive parameters for the designing of optical devices, i.e., EM field, source transformation, multi-beams, field-concentrators and rotators, isotropic emissions, and invisibility cloaks [5–35] etc. Similarly, directive antennas have been introduced employing the TO to achieve high-directivity from the smaller-sized antenna [36–41]. Overall, the TO technique is useful in constructing the conceptual devices, whereas the hurdle of complex constitutive parameters can be minimized by achieving homogeneous-materials using linear coordinate transformation method [19, 39, 42].

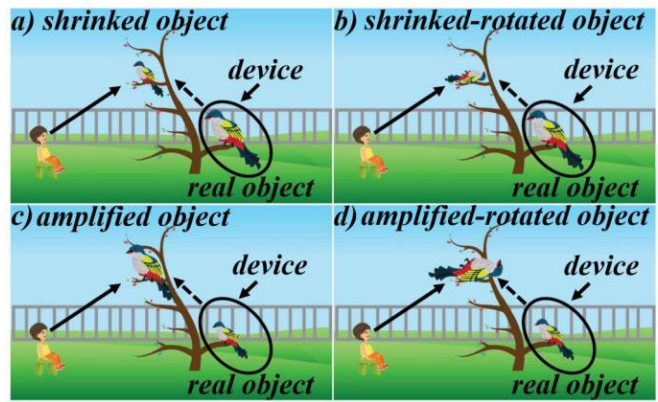

**Fig. 1.** The simple schematic example for four types of shrinking and amplifying devices. **(a)** Shrinking-shifting medium layer cause a big bird appears like a smaller bird placed at different location. **(b)** Shrinking-reshaping medium layer cause the big bird looks like simultaneously smaller and rotated bird placed at different location. **(c)** Amplifying-shifting medium layer cause a small bird appears like a bigger bird placed at different location. **(d)** Amplifying-reshaping medium layer cause a small bird looks like simultaneously bigger and rotated bird placed at different location.

The human's perception of outer environment and world is due to the brain judgment based on the information of five-senses. Among these senses, visual-sense is important in making the best perception. Thus, the optical-illusion phenomena refer in which the visual-system perceives illusory-objects with different color, shape, orientation, and motion from the real object in real space. TO tool is helpful to design illusion devices that fool the viewer and (or) detector in order to make wrong judgment [42–52]. Such illusion devices transform the real object into the object chosen for illusion. Similarly, a bi-functional illusion device was investigated in which the device performance is dependent on the shifting region that exists outside of the physical space boundaries. In this regard, the actual object was limited to stay outside the device boundaries [52]. Up to now, various illusion devices have been designed for the different functionalities including multi-physical fields (acoustics, thermal-dynamics, and electrostatics [53–55]), imaging devices, superscatterers, illusion devices for passive and active scatterer, super absorbers, super resolution, overlapping illusion, shrinking devices, and amplifying devices [56–63] etc.

To best of our knowledge, the first design of illusion device was superscatterers based on complementary medium to amplify the small object into bigger one [61]. However, this device contains double-negative material properties that make it difficult for practical implementation. To remove these flaw, other illusion devices made by positive materials came to exist [44,46,47]. On the other hand, shrinking device was investigated to make big object virtually appears as a small object but with different material properties [44]. Although, the negative material-property was successfully removed, but anisotropic and inhomogeneous properties remains a challenge for fabrication. In this regard, symmetric shrinking device was introduced to remove the anisotropic and inhomogeneous property [60]. Unlike shrinking device, an amplifying device has also been investigated to make a small object virtually appears like a bigger one [59]. Moreover, another homogeneous based illusion device has also been reported that exhibits both shifted and transformation effect simultaneously [57].

However, it can be noticed that the above-mentioned devices either possess inhomogeneous and anisotropic properties or compose of specific symmetric geometries that cause hurdle in the flexibility of device designing. Similarly, previously designed shrinking and amplifying devices are single-tasking devices with lack of multiple functionalities to perform simultaneously [42–63]. It seems like in previous designs, the object's shrinking and amplifying has been referred as simply changing the object's size and materials etc. Meanwhile, sometimes we need object protection by virtually shift the actual location of object. Thus, a need still exists to design a bi-functional shrinking and amplifying devices for re-sizing, location- and- position shifting.

Here, we investigate the homogeneous bi-functional shrinking device that performs re-sizing and location-shifting properties that are independent on the shifting-distance between the shifting region and physical space. Firstly, by properly geometric divisions of the shifting-region and physical space, the shrinking-shifting device is investigated that behaves like a traditional-concentrator with the absence of any object [11]. It is worth mentioning that when either the active scatterer or passive scatterer is wrapped inside the core-region of proposed device, the wrapped-object behaves optically similar to a smaller-sized scatterer that is shifted at different location. In Fig. 1(a), a simple schematic example of proposed shrinking device, in which a bird wrapped by a shrinking medium layer to make it appears like a smaller bird at different location. Most importantly, the proposed device contains positive materials that may open doors for designing devices in other physical fields, i.e., acoustics, thermodynamics, and electrostatics, etc. Secondly, a distinct mapping method used for traditional-rotator approach is employed to design a shrinking-reshaping device that gives impression that the scatterer is at different location with different position, can be seen in Fig. 1(b). Similarly, the same distinct mapping method used for concentrator (Fig. 1(a)) and rotator (Fig. 1(b)) approach is employed to design amplify-shifting and amplify-reshaping devices respectively, to pursue the proposed concept of making small object appear like bigger one. In Fig. 1(c) and Fig. 1(d), are the simple schematic examples of proposed amplifying devices, in which a bird wrapped by an amplifying medium layer to make it appears like a bigger bird at different location and with different position, respectively. The full wave finite element method (COMSOL) is used to demonstrate and validate the proposed concept. Hence, the simulation results are in good match with the expected behavior of proposed interesting phenomena. We remark that the proposed idea will find useful applications in protecting objects either by showing more strength or showing low strength from different location than its actual shape, strength and location.

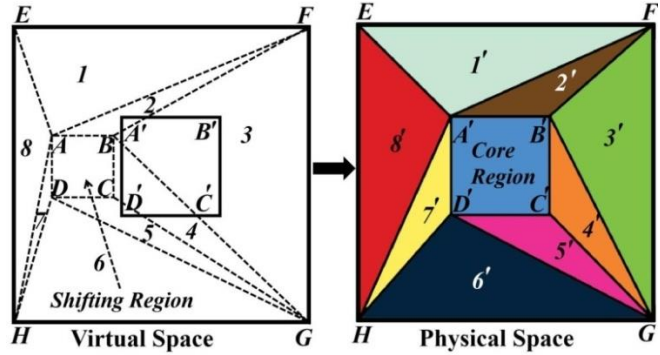

**Fig. 2.** The schematic diagram of the proposed shrinking-shifting device. The concentrator-mapping style of the dashed line regions in virtual space, denoted as $1, 2, 3, 4...8$, is accordingly mapped to $1', 2', 3', 4'...8'$. Whereas, the shifting region at origin $(-0.25, 0)$ is shifted towards the core region at origin $(0, 0)$.

## 2. MATERIALS AND METHODS

To start with, a schematic diagram illustrating the prescription of the proposed resizing and reshaping device is given in Fig. 2. The regions inside the dashed lines are used for virtual space and after mapping, these regions are transformed into physical space. The aim of this mapping style is to shift the shifting region $(ABCD)$ to a new region $(A'B'C'D')$, where the origin point of shifting-region is $(-0.25, 0)$, and the origin point of core-region is $(0, 0)$. Hereafter, it is well known that the simplicity of the material parameters makes it easier for practical implementation.

To avoid the material complexity, we divide the shifting region of $N$ regions into $N$ triangular-shaped regions, in accordance with the physical space. In this regard, the virtual space of eight different regions is mapped to the same number of regions, but at different locations, to obtain the desired goal. The imaginary space of the total eight regions is donated by $1, 2, 3, 4...8$. Firstly, the shifting region $(ABCD)$ in the virtual space $(x, y, z)$ is shifted towards $(A'B'C'D')$ in the physical space $(x', y', z')$. After that, as an example, region 1 of $\Delta EFA$ and region 2 of $\Delta FAB$ in the virtual space $(x, y, z)$ are mapped into region $\Delta EFA'$ labeled as $1'$ and $\Delta FA'B'$ labeled as $2'$, in the physical space $(x', y', z')$, respectively. Similarly, the other regions in imaginary space, denoted as $2, 3, 4...8$, are accordingly mapped into regions $2', 3', 4'...8'$ to design the proposed device, where $\varepsilon = \mu = 1$ represent the material parameters of free space and shifting-region, while $\varepsilon_{core}, \mu_{core}$ shows the constitutive parameters of core-region. Thus, the whole transformation procedure is fully devoted to obtain bi-functional effects of both shrinking and position shifting. Meanwhile, the constitutive parameters of each region can be found with the help of the following equations [19]:

$$\varepsilon' = \frac{J \varepsilon_0 J^T}{\det J}$$
$$\mu' = \frac{J \mu_0 J^T}{\det J}$$
(1)

where J is the Jacobian tensor, and for the triangular shape only in 2D transformation, the Jacobian matrix can be found as [19]:

$$J = \begin{bmatrix} x_3' - x_1' & x_2' - x_1' & 0 \\ y_3' - y_1' & y_2' - y_1' & 0 \\ 0 & 0 & 1 \end{bmatrix} \begin{bmatrix} x_3 - x_1 & x_2 - x_1 & 0 \\ y_3 - y_1 & y_2 - y_1 & 0 \\ 0 & 0 & 1 \end{bmatrix}^{-1} \quad (2)$$

In the following, a full wave simulation of the finite element method (COMSOL) is employed in a two-dimensional scenario to validate the effectiveness and correctness of the proposed bi-functional devices.

## 3. RESULTS AND DISCUSSIONS

In this paper, we claim that the mapping style determines the properties of the proposed device, in which, the device behaves as a shrinking-shifting device. It is worth noting that for shrinking-effect, the size of the shifting-region $[(ABCD)]$ is smaller than the core-region in the physical space $[(A'B'C'D')]$. The proposed device is applicable to both transverse electric (TE) and transverse magnetic (TM) mode. Here, as an example, the TE mode is adopted at the frequency of 1 GHz.

*A. Shrinking and Shifting*

To make the design methodology easy to understand, the coordinate values of each point for both imaginary and physical space in SI units are given as following: $E(-0.4, 0.4)$, $F(0.4, 0.4)$, $G(0.4, -0.4)$, $H(-0.4, -0.4)$, $A(-0.325, 0.075)$, $B(-0.175, 0.075)$ $C(-0.175, -0.075)$, $D(-0.325, -0.075)$, $A'(-0.15, 0.15)$, $B'(0.15, 0.15)$, $C'(0.15, -0.15)$, and $D'(-0.15, -0.15)$. When region 1 and region 2 are mapped into region 1' and region 2', respectively (Fig. 2), the coordinate parameters of each point for each region are already known. Thus, by taking advantage of Eqs. (1–2), the material parameters will become: $\varepsilon_1 = \mu_1 = \begin{bmatrix} 1.6769 & -0.5385 & 0 \\ -0.5385 & 0.7692 & 0 \\ 0 & 0 & 1.3 \end{bmatrix}$, and $\varepsilon_2 = \mu_2 = \begin{bmatrix} 7.5846 & -1.3846 & 0 \\ -1.3846 & 0.3846 & 0 \\ 0 & 0 & 0.65 \end{bmatrix}$. The same method is applied to find the constitutive parameters of the remaining regions, such as: $\varepsilon_3 = \mu_3 = \begin{bmatrix} 0.4348 & -0.1304 & 0 \\ -0.1304 & 2.3391 & 0 \\ 0 & 0 & 2.3 \end{bmatrix}$, $\varepsilon_4 = \mu_4 = \begin{bmatrix} 0.2174 & 0.3478 & 0 \\ 0.3478 & 5.1565 & 0 \\ 0 & 0 & 1.15 \end{bmatrix}$, $\varepsilon_5 = \mu_5 = \begin{bmatrix} 7.5846 & 1.3846 & 0 \\ 1.3846 & 0.3846 & 0 \\ 0 & 0 & 0.65 \end{bmatrix}$, $\varepsilon_6 = \mu_6 = \begin{bmatrix} 1.6769 & 0.5385 & 0 \\ 0.5385 & 0.7692 & 0 \\ 0 & 0 & 1.3 \end{bmatrix}$, $\varepsilon_7 = \mu_7 = \begin{bmatrix} 1.6667 & -2.6667 & 0 \\ -2.6667 & 4.8667 & 0 \\ 0 & 0 & 0.15 \end{bmatrix}$, and $\varepsilon_8 = \mu_8 = \begin{bmatrix} 3.3333 & 1 & 0 \\ 1 & 0.6 & 0 \\ 0 & 0 & 0.3 \end{bmatrix}$. In addition, as described earlier, the shifting-region $(ABCD)$ is shifted towards $(A'B'C'D')$ in physical space and for simplicity; we further divide the shifting region into two triangles to fulfill the criteria of equation (2). Hence, the material properties of this core region will become: $\varepsilon_{core} = 0.25$, $\mu_{core} = 1$.

The specific aim of this paper is still questioned to determine whether it is possible to provide a strategic tool in favor of a shrinking for any type of scatterer. In this context, examples of different types of scatterer are considered in the following. In the first case, a passive scatterer is manipulated to achieve the required field pattern of a larger object to give the impression that the object is smaller in size and placed at a different position. Similarly, in the second example, shrinking and shifting of the active scatterer is achieved by manipulating the radiation pattern of the antenna. Thus, in the presence of any scatterers wrapped inside the core region, the proposed device (in Fig. 2) will behave like a bi-functional shrinking device. Most importantly, the proposed scheme is independent of anti-mirror object [10].

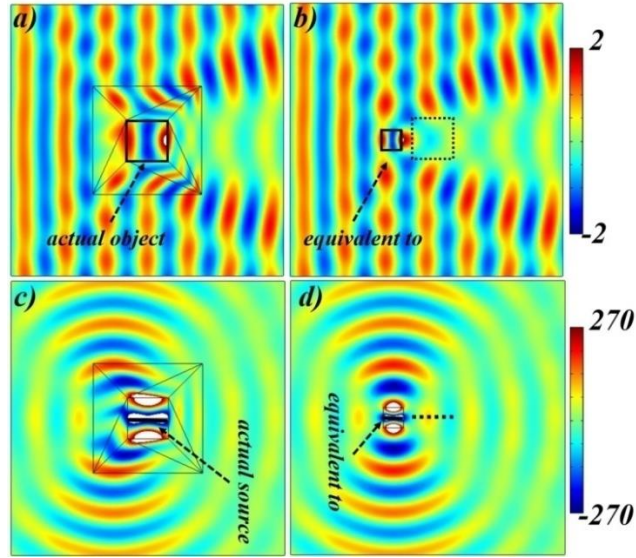

**Fig. 3.** Simulation results of both passive and active scatterer under shrinking-shifting device to achieve shrinking and location shifting. **(a)** The z-directed electric-field's scattering behavior of big-sized dielectric object of $\varepsilon = 5$ and is $\mu = 1$ is placed in the core-region of the proposed device that is optically equivalent to a smaller-sized object in a different position, as shown in **(b)**. **(c)** $E_z$ distributions of a big-sized linear source with electric-field intensity of $1V$ under the proposed device is manipulated to achieve shrinking and location shifting. This gives the impression of the smaller-sized source's existence in a different position **(d)**.

For detailed discussion, Fig. 3 describes the electric-field ($E_z$) distribution when the scatterer is kept in the core region which is generating waves inside the core region. Fig. 3(a) expresses an object of size $A'B'C'D'$ having characteristics of $\varepsilon = 5$ and $\mu = 1$ is wrapped in the proposed device of Fig. 2 and interacts with the plane waves. Thus, the detector can detect that the field pattern has been modified in such a design which is optically equivalent to Fig. 3(b). Whereas, the object size $A'B'C'D'$ (in Fig. 3(a)) is comparatively bigger than $ABCD$ (in Fig. 3(b)) and the location is also different. Similarly, the verification of the proposed design is further implemented on active scatterer with the electric-field having intensity of $1V$. Figs. 3(c)–3(d) represents the simulation results where the radiation patterns of the line source in Fig. 3(c) is placed at $(-0.15, 0), (0.15, 0)$ to achieve shrinking and position shifting by the proposed device. This is optically equivalent to Fig. 3(d) in which the source position is at $(-0.325, 0), (-0.175, 0)$.

## B. Shrinking, Reshaping and Shifting

As discussed above, the shrinking and position-shifting effect of the proposed device is effective when the concentrator-mapping style [11] is employed among shifting-region and core-region. Now, the question arises that what happens to the shrinking feature when mapping-style is changed to rotational-mapping between shifting-region and core-region?

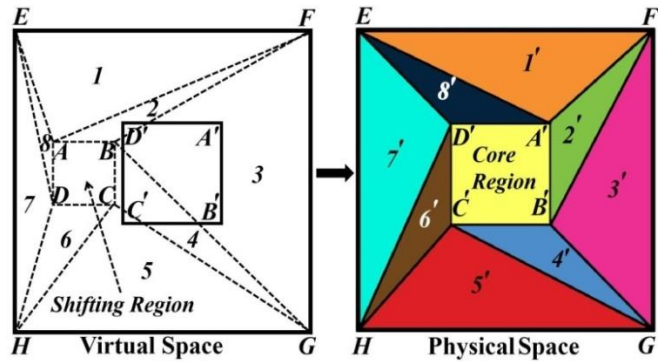

**Fig. 4.** The schematic diagram of the proposed shrinking-reshaping device. The design methodology is similar to that in Fig. 2 except the mapping style. The mapping of the dashed line regions in virtual space denoted as $1, 2, 3, 4...8$ are accordingly mapped to $1', 2', 3', 4'...8'$.

To answer this question, the idea presented in Fig. 2 is further extended as that we can manipulate the field patterns in such a way causing object's shape position and location shifting. A schematic diagram of the shrinking-reshaping and- shifting device has been demonstrated in Fig. 4, in which a shifting-region in virtual space is presented by dashed lines. Therefore, the virtual space is transformed into the physical space after recalling Fig. 2. The coordinate values of each point for both the imaginary and physical space in SI units are given as the following: $A'(0.15, 0.15)$, $B'(0.15, -0.15)$, $C'(-0.15, -0.15)$, and $D'(-0.15, 0.15)$, while the remaining coordinate values of other points are similar, as described in Fig. 2. Similarly, the material parameters of the physical space can be calculated using Eqs. 1–2, thus, the material parameters will become:

$$\varepsilon_1 = \mu_1 = \begin{bmatrix} 4.0769 & -1.4615 & 0 \\ -1.4615 & 0.7692 & 0 \\ 0 & 0 & 1.3 \end{bmatrix}, \quad \varepsilon_2 = \mu_2 = \begin{bmatrix} 0.3846 & 2.6154 & 0 \\ 2.6154 & 20.3846 & 0 \\ 0 & 0 & 0.65 \end{bmatrix}, \quad \varepsilon_3 = \mu_3 = \begin{bmatrix} 0.4348 & 0.3913 & 0 \\ 0.3913 & 2.6522 & 0 \\ 0 & 0 & 2.3 \end{bmatrix},$$

$$\varepsilon_4 = \mu_4 = \begin{bmatrix} 9.6087 & -1.0435 & 0 \\ -1.0435 & 0.2174 & 0 \\ 0 & 0 & 1.15 \end{bmatrix}, \quad \varepsilon_5 = \mu_5 = \begin{bmatrix} 1.3077 & 0.0769 & 0 \\ 0.0769 & 0.7692 & 0 \\ 0 & 0 & 1.3 \end{bmatrix}, \quad \varepsilon_6 = \mu_6 = \begin{bmatrix} 0.3846 & 1.0769 & 0 \\ 1.0769 & 5.6154 & 0 \\ 0 & 0 & 0.65 \end{bmatrix},$$

$$\varepsilon_7 = \mu_7 = \begin{bmatrix} 3.3333 & 3 & 0 \\ 3 & 3 & 0 \\ 0 & 0 & 0.3 \end{bmatrix}, \text{ and } \varepsilon_8 = \mu_8 = \begin{bmatrix} 39 & -8 & 0 \\ -8 & 1.6667 & 0 \\ 0 & 0 & 0.15 \end{bmatrix}.$$

In addition, as described earlier, the shifting region $(ABCD)$ is shifted towards $(A'B'C'D')$ in the physical space, and the material properties of this core region will become: $\varepsilon_{core} = 0.25$, $\mu_{core} = 1$.

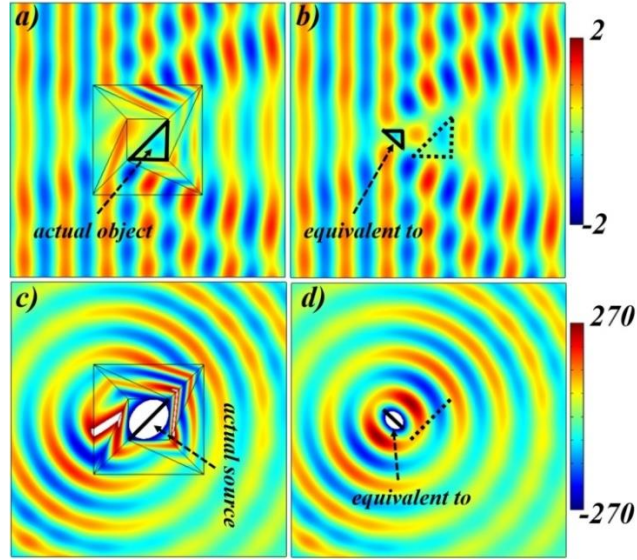

**Fig. 5.** Simulation results of both passive and active scatterer under shrinking-reshaping device to achieve shrinking, reshaping and location shifting. **(a)** The z-directed electric-field's scattering behavior of big-sized dielectric object of $\varepsilon = 5$ and is $\mu = 1$ is placed in the core region of the proposed device that is optically equivalent to a smaller-sized object in a different position and location, as shown in **(b)**. **(c)** $E_z$ distributions of a big-sized line source with electric-field intensity of 1V under the proposed device is manipulated to achieve shrinking and location shifting. This gives the impression of the smaller-sized source's existence in a different position and location **(d)**.

Here, we want to shrink-shift and reshape the field pattern of object (size of $\Delta A'B'C'$) in Fig. 5(a). The viewer can observe the modified field pattern that is optically equivalent to the object of size $\Delta ABC$ (Fig. 5(b)). Whereas, the shape and size of object $\Delta ABC$ is smaller than $\Delta A'B'C'$ and the location of object is also different. A line source in Fig. 5(c) is located at coordinates $\overline{A'C'}$ in the proposed device which changes the location while the field pattern and location (in Fig. 5(c)) is optically same as that of Fig. 5(d). However, the coordinates' values of the located source in Fig. 5(d) are $\overline{AC}$. These results provide more authenticity to justify the proposed concept of the shrinking-shifting and reshaping.

## C. Amplifying and Shifting

All the positive results for the above-described sections represent that we can shrink-shift and reshaping of the scatterer, simultaneously. In some scenarios, we require amplification to virtually appear the smaller size object into bigger one. In this regard, we present a different rotational-mapping approach to design an amplifying-shifting device. In this regard, the mapping methodology is demonstrated in Fig. 6.

There is no doubt that the all the mapping style and coordinate values of each point is similar to that of Fig. 2, except the different origin points of both shifting-region ($ABCD$) and core-region ($A'B'C'D'$) in the physical space. Here, the origin point of shifting-region is $(0,0)$, and $(-0.25,0)$ is the origin of core-region. The coordinate values for both core-region and shifting-region in SI units are given as following: $A'(-0.325, 0.075)$, $B'(-0.175, 0.075)$ $C'(-0.175, -0.075)$, $D'(-0.325, -0.075)$, $A(-0.15, 0.15)$, $B(0.15, 0.15)$, $C(0.15, -0.15)$, and $D(-0.15, -0.15)$. In addition, the material parameters can be calculated by recalling Fig. 2, which will become:

$$\varepsilon_1 = \mu_1 = \begin{bmatrix} 1.1462 & 0.7 & 0 \\ 0.7 & 1.3 & 0 \\ 0 & 0 & 0.7692 \end{bmatrix}, \varepsilon_2 = \mu_2 = \begin{bmatrix} 5.3692 & 3.6 & 0 \\ 3.6 & 2.6 & 0 \\ 0 & 0 & 1.5385 \end{bmatrix}, \varepsilon_3 = \mu_3 = \begin{bmatrix} 2.3 & 0.3 & 0 \\ 0.3 & 0.4739 & 0 \\ 0 & 0 & 0.4348 \end{bmatrix},$$

$$\varepsilon_4 = \mu_4 = \begin{bmatrix} 4.6 & -1.6 & 0 \\ -1.6 & 0.7739 & 0 \\ 0 & 0 & 0.8696 \end{bmatrix}, \varepsilon_5 = \mu_5 = \begin{bmatrix} 2.0692 & -1.3 & 0 \\ -1.3 & 1.3 & 0 \\ 0 & 0 & 0.7692 \end{bmatrix}, \varepsilon_6 = \mu_6 = \begin{bmatrix} 0.4462 & -0.4 & 0 \\ -0.4 & 2.6 & 0 \\ 0 & 0 & 1.5385 \end{bmatrix},$$

$$\varepsilon_7 = \mu_7 = \begin{bmatrix} 0.6 & 1.6 & 0 \\ 1.6 & 5.9333 & 0 \\ 0 & 0 & 6.6667 \end{bmatrix}, \text{ and } \varepsilon_8 = \mu_8 = \begin{bmatrix} 0.3 & -0.3 & 0 \\ -0.3 & 3.6333 & 0 \\ 0 & 0 & 3.3333 \end{bmatrix}.$$ Similarly, after recalling Fig. 2, the material properties of this core region will become: $\varepsilon_{core} = 4, \mu_{core} = 1$.

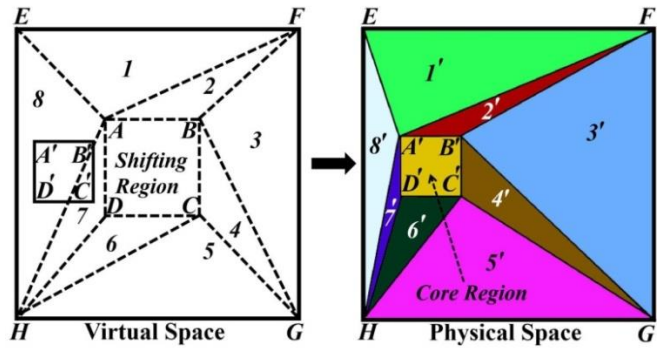

**Fig. 6.** The schematic diagram of the proposed amplify-shifting device. The mapping methodology is similar to that in Fig. 2. Here, the origin of shifting region is $(0,0)$ and the origin of core-region is $(-0.25,0)$. The mapping of the dashed line regions in virtual space denoted as $1,2,3,4...8$ are accordingly mapped to $1',2',3',4'...8'$.

The electric-field ($E_z$) distribution can be seen in Fig. 7(a), in which the object of size $A'B'C'D'$ having properties of $\varepsilon = 5$ and $\mu = 1$ is placed inside the core-region of proposed device. Hence, the amplification-shifting has been done as the object size $A'B'C'D'$ in Fig. 7(a) is smaller, and amplified to bigger object $ABCD$ with location-shifting (Fig. 7(b)). Moreover, the material characteristics of $ABCD$ are $\varepsilon = 5$ and $\mu = 1$. Furthermore, we implement for an active scatterer having electric field intensity of $1V$. In this situation, line source is placed at $(-0.325, 0), (-0.175, 0)$ (Fig. 7(c)) and its radiation pattern is adjusted in such a way that it can achieve increasing size with location-shifting by surrounding it with the proposed device. This is optically same as in Fig. 7(d), in which source located at $(-0.15, 0), (0.15, 0)$.

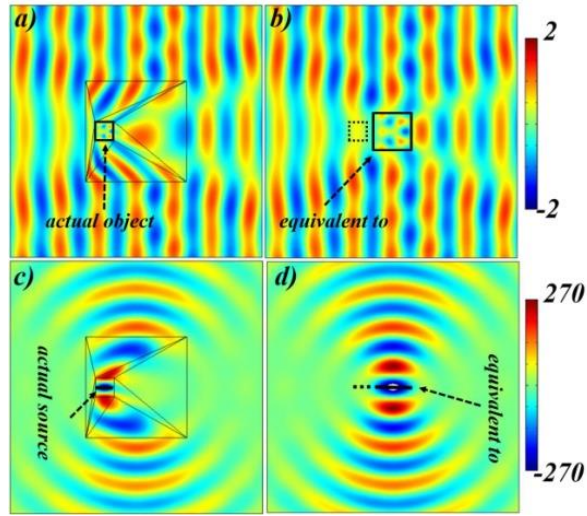

**Fig. 7.** Simulation results of both passive and active scatterer under amplify-shifting device to achieve amplifying and location shifting. **(a)** The z-directed electric-field's scattering behavior of small-sized dielectric object of $\varepsilon = 5$ and is $\mu = 1$ is placed in the core region of the proposed device that is optically equivalent to a bigger object at different location **(b)**. **(c)** $E_z$ distribution of small-sized linear source with electric-field intensity of $1V$ under the proposed device gives impression of the bigger source's existence at different location **(d)**.

## D. Amplifying, Shifting and Reshaping

From Section C, it has been concluded that the amplifying and location-shifting effect of the proposed device is effective using concentrator-mapping style [11]. Now the question is, if we adopt the rotational-mapping between shifting region and core region, then what will happen?

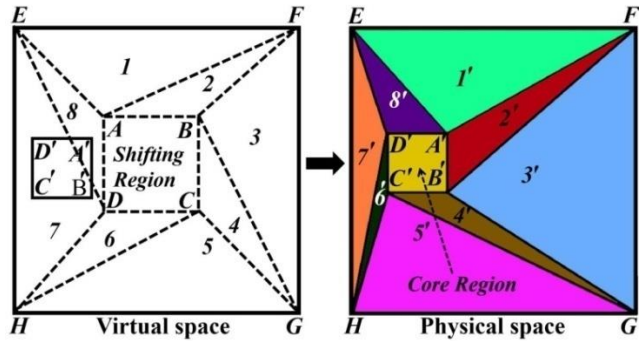

**Fig. 8.** The schematic diagram of the proposed amplify-reshaping device. The mapping methodology is similar to that of Fig. 4 with the coordinate values of Fig. 6.

To answer this question, the idea presented in Fig. 6 is further extended as that we can manipulate the field patterns in such a way causing amplify-shifting and object rotation. A schematic diagram of the amplify-shifting and reshaping device has been demonstrated in Fig. 8, in which a rotation and amplification mapping style in the virtual space is shown by dashed lines, next to the physical space. The virtual space after mapping is transformed into the physical space after recalling Fig. 4. The coordinate values of each point for both the imaginary and physical space in SI units are given as the following: $A'(-0.175, 0.075)$, $B'(-0.175, -0.075)$, $C'(-0.325, -0.075)$, and $D'(-0.325, 0.075)$, while the remaining coordinate values of other points are similar, as described in Fig. 6. Similarly, the material parameters of the physical space can be calculated using Eqs. (1)–(2). Thus, the material parameters will become:

$$\varepsilon_1 = \mu_1 = \begin{bmatrix} 0.7769 & 0.1 & 0 \\ 0.1 & 1.3 & 0 \\ 0 & 0 & 0.7692 \end{bmatrix}, \quad \varepsilon_2 = \mu_2 = \begin{bmatrix} 4.6 & 4.8 & 0 \\ 4.8 & 5.2261 & 0 \\ 0 & 0 & 0.8696 \end{bmatrix}, \quad \varepsilon_3 = \mu_3 = \begin{bmatrix} 2.3 & 0.9 & 0 \\ 0.9 & 0.7870 & 0 \\ 0 & 0 & 0.4348 \end{bmatrix},$$

$$\varepsilon_4 = \mu_4 = \begin{bmatrix} 18.1692 & -6.8 & 0 \\ -6.8 & 2.6 & 0 \\ 0 & 0 & 1.5385 \end{bmatrix}, \quad \varepsilon_5 = \mu_5 = \begin{bmatrix} 3.5462 & -1.9 & 0 \\ -1.9 & 1.3 & 0 \\ 0 & 0 & 0.7692 \end{bmatrix}, \quad \varepsilon_6 = \mu_6 = \begin{bmatrix} 0.6 & 4.8 & 0 \\ 4.8 & 40.0667 & 0 \\ 0 & 0 & 6.6667 \end{bmatrix},$$

$$\varepsilon_7 = \mu_7 = \begin{bmatrix} 0.3 & 0.9 & 0 \\ 0.9 & 6.0333 & 0 \\ 0 & 0 & 3.3333 \end{bmatrix}, \text{ and } \varepsilon_8 = \mu_8 = \begin{bmatrix} 3.4 & -2.8 & 0 \\ -2.8 & 2.6 & 0 \\ 0 & 0 & 1.5385 \end{bmatrix}.$$ Similarly, after recalling Fig. 4, the material properties of this core region will become: $\varepsilon_{core} = 4, \mu_{core} = 1$.

In Fig. 9(a), we intend to amplify and reshape the field pattern of abject size $\Delta A'B'C'$, so that the observer can see the field pattern optically equivalent to Fig. 9(b) with the object size $\Delta ABC$. Whereas, the size of object $\Delta ABC$ is observed as bigger than $\Delta A'B'C'$, also with different shape and location. Similarly, a line source of coordinates $\overline{A}'\overline{C}'$ is placed inside the proposed device (Fig. 9(c)), which behaves like Fig. 9(d). However, $\overline{AC}$ are the coordinate values of the located source (Fig. 9(d)).

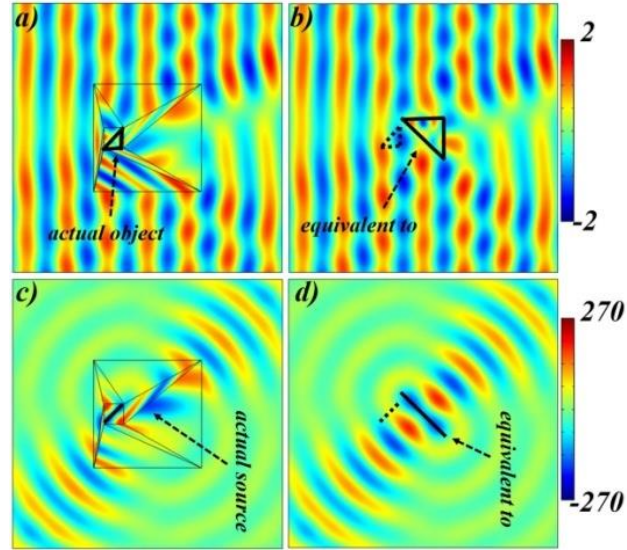

**Fig. 9.** Simulation results of both passive and active scatterer under amplifying-shifting and reshaping device to achieve amplification, reshaping and location shifting. **(a)** The z-directed electric-field's scattering behavior of small-sized dielectric object of $\varepsilon = 5$ and is $\mu = 1$ is placed in the core region of the proposed device that is optically equivalent to bigger object at different location with different shape **(b)**. **(c)** $E_z$ distributions of small-sized linear source with electric-field intensity of $1V$ under the proposed device is manipulated that gives the impression of bigger source's existence with different position and location **(d)**.

Four different examples are discussed in this paper in which the shifting region has been smoothly moved from and to core-region, and different mapping styles have been applied. It is observed that the constitutive parameters of proposed devices are positive, homogeneous and anisotropic values.

## 4. CONCLUSION

A bi-functional shrinking device has been proposed and constructed based on a linear coordinate transformation method that exhibits resizing and position-shifting features depending on the concentrator-mapping style between the shifting-region and core-region. Four different examples are presented, constitutive parameters are prescribed, and the simulation results validate the correctness of the proposed designs. The relative position of the shifting-region and core-region, and its effect on the resizing and shifting performances are discussed. When the shifting region moves from the origin (0,0) in the specific concentrator-mapping style, the device behaves as shrinking-shifting device. When the mapping style is rotational, the proposed device will act as a shrinking-reshaping device. Similarly, when the shifting region moves to the origin (0,0) in the specific concentrator approach and rotational way, the device will act as amplifying-shifting and amplifying-reshaping, respectively. The proposed devices possess positive, homogeneous and anisotropic values materials. We are confident that the proposed concept will open avenues for potential applications in existing shrinking, amplifying, and cloaking technologies.

**FUNDING.** This work was supported by the National Key Research and Development Program of China (2017YFA0700201, 2017YFA0700202, 2017YFA0700203), the National Natural Science Foundation of China (61631007, 61571117, 61501112, 61501117, 61522106, 61731010, 61735010, 61722106, 61701107, and 61701108), and the 111 Project (111-2-05). H. A. Madni acknowledges the support of Department of Computer Engineering, Khwaja Fareed University of Engineering and Information Technology, Rahim Yar Khan 64200, Pakistan.

**ACKNOWLEDGMENT.** H. A. Madni designed the devices and carried out the simulations. M. M. Bilal, F. Jaleel, A. Sohaib and W. X. Jiang analyzed the data and interpreted the results. H. A. Madni drafted the manuscript with the input from the others. W. X. Jiang supervised the project. The authors would like to thank the Editor and the anonymous reviewers for their insightful comments and constructive suggestions that certainly improved the quality of this paper.

**Conflicts of Interest:** The authors declare no conflict of interest.

## REFERENCES

[1] J. B.Pendry, D.Schurig, and D. R.Smith, "Controlling electromagnetic fields," Science **312**, 1780–2 (2006).
[2] H. A.Madni, B. Zheng, Y. Yang, H. Wang, X. Zhang, W.Yin, E.Li, and H. Chen, "Non-contact radio frequency shielding and wave guiding by multi-folded transformation optics method," Sci. Rep. **6**, 36846 (2016).
[3] T. J.Cui, "Microwave metamaterials," Natl Sci. Rev. **5**, 134–6 (2018).
[4] T. J.Cui, "Microwave metamaterials—from passive to digital and programmable controls of electromagnetic waves," J. Opt. **19**, 084004 (2017).
[5] D.Schurig, J.J.Mock, B. J.Justice, S. A.Cummer, J. B.Pendry, A. F.Starr, and D. R.Smith, "Metamaterial electromagnetic cloak at microwave frequencies," Science**314**, 977–80 (2006).
[6] J.Li, and J. B.Pendry, "Hiding under the carpet: a new strategy for cloaking," Phys. Rev. Lett. **101**, 203901(2008).
[7] U.Leonhardt, and T.Tyc, "Broadband invisibility by noneuclidean cloaking," Science **323**,110–2 (2009).
[8] B.Zhang, Y.Luo, X.Liu, and G.Barbastathis, "Macroscopic invisibility cloak for visible light," Phys. Rev. Lett. **106,**033901 (2011).
[9] B.Zheng, H, A.Madni, R.Hao, X.Zhang, X.Liu, E.Li, and H.Chen, "Concealing arbitrary objects remotely with multi-folded transformation optics," Light-Sci. Appl. **5**, e16177(2016 ).
[10] Y.Lai, H.Chen, Z. Q.Zhang, and C. T.Chan, "Complementary media invisibility cloak that cloaks objects at a distance outside the cloaking shell," Phys. Rev. Lett. **102**, 093901 (2009).
[11] Y.Luo, H.Chen, J.Zhang, L.Ran, and J. A.Kong, "Design and analytical full-wave validation of the invisibility cloaks, concentrators, and field rotators created with a general class of transformations," Phys. Rev. B. **77**, 125127 (2008).
[12] H.Chen, B.Hou, S.Chen, X.Ao, W.Wen, and C. T.Chan, "Design and experimental realization of a broadband transformation media field rotator at microwave frequencies,"Phys. Rev. Lett. **102**, 183903 (2009).
[13] H.Chen, and C. T.Chan, "Transformation media that rotate electromagnetic fields," Appl. Phys.Lett. **90**, 241105 (2007).
[14] T.Zhai, Y.Zhou, J.Zhou, and D.Liu, "Polarization splitter and polarization rotator designs based on transformation optics," Opt. Express **16**, 18731–8 (2008).
[15] M.Rahm, D.Schurig, D. A.Roberts, S. A.Cummer, D. R.Smith, and J. B.Pendry, "Design of electromagnetic cloaks and concentrators using form-invariant coordinate transformations of Maxwell's equations," Photon. Nanostruct. Fundam.Appl. **6**, 87 (2008).
[16] A. D.Yaghjian, and S.Maci, "Alternative derivation of electromagnetic cloaks and concentrators," New J. Phys. **10**, 115022 (2008).
[17] C.Navau, J. P.Camps, and A.Sanchez, "Magnetic energy harvesting and concentration at a distance by transformation optics," Phys. Rev. Lett. **109**, 263903 (2012).
[18] B.Bian, S.Liu, S.Wang, X.Kong, Y.Guo, X.Zhao, B.Ma and C.Chen, "Cylindrical optimized nonmagnetic concentrator with minimized scattering," Opt. Express **21**, A231–40 (2013).
[19] H. A.Madni, K.Hussain, W. X.Jiang, S.Liu, A.Aziz, S.Iqbal, A.Mahboob, and T. J.Cui, "A novel EM concentrator with open-concentrator region based on multi-folded transformation optics," Sci. Rep. **8,** 9641 (2018).

[20] G. X.Yu, W. X.Jiang, X. Y.Zhou, and T. J.Cui, "Non-rotationally invariant invisibility cloaks and concentrators of EM waves," Eur.Phys. J. Appl.Phys. **44**, 181–5 (2008).

[21] L.Lin, W.Wang, C.Du, and X.Luo, "A cone-shaped concentrator with varying performances of concentrating," Opt. Express **16**, 6809–14 (2008).

[22] J.Yang, M.Huang, C.Yang, Z.Xiao, and J.Peng, "Metamaterial electromagnetic concentrators with arbitrary geometries," Opt. Express **17,** 19656–61 (2009).

[23] W.Li, J.Guan, and W.Wang, "Homogeneous-materials constructed electromagnetic field concentrators with adjustable concentrating ratio," J. Phys. D: Appl. Phys. **44**,125401 (2011).

[24] K.Zhang, Q.Wu, J.H.Fu, and L.W.Li, "Cylindrical electromagnetic concentrator with only axial constitutive parameter spatially variant," J. Opt. Soc. Am. B. **28**, 1573–7 (2011).

[25] D.H.Kwon, "Transformation electromagnetic design of an embedded monopole in a ground recess for conformal applications," IEEE Antenn. Wirel. Pr.**9**, 432–5 (2010).

[26] B.I.Popa, J.Allen, and S. A.Cummer, "Conformal array design with transformation electromagnetics," Appl. Phys.Lett. **94**, 244102 (2009).

[27] F.Kong, B.I.Wu, J. A.Kong, J.Huangfu, S,Xi, and H.Chen, "Planar focusing antenna design by using coordinate transformation technology," Appl. Phys. Lett. **91**, 253509 (2007).

[28] W.Lu, Z.Lin, H.Chen, and C. T.Chan, "Transformation media based super focusing antenna," J. Phys. D: Appl. Phys. **42**, 212002 (2009).

[29] J.Zhang, Y.Luo, H.Chen, and B.I.Wu, "Manipulating the directivity of antennas with metamaterial," Opt. Express **16**, 10966 (2008).

[30] Y.Luo, J.Zhang, L.Ran, H.Chen, and J. A.Kong, "New concept conformal antennas utilizing metamaterial and transformation optics," IEEE Antenn. Wirel. Pr. **7**, 509–12 (2008).

[31] Z. H.Jiang, M. D.Gregory, and D. H.Werner, "Experimental demonstration of a broadband transformation optics lens for highly directive multi-beam emission," Phys. Rev. B. **84**, 165111 (2011).

[32] Z. H.Jiang, M. D.Gregory, and D. H.Werner, "Broadband high directivity multi beam emission through transformation optics-enabled metamaterial lenses," IEEE Trans. AntennasPropag. **60**, 5063 (2012).

[33] P.H.Tichit, S. N.Burokur, and A.de Lustrac, "Spiral-like multi-beam emission via transformation electromagnetics," J. Appl.Phys. **115**, 024901 (2014).

[34] P.H.Tichit, S. N.Burokur, and A.de Lustrac, "Transformation media producing quasi-perfect isotropic Emission," Opt. Express.**19**, 20551 (2011).

[35] P.H.Tichit, S. N.Burokur, C.W.Qiu, and A.de Lustrac, "Experimental verification of isotropic radiation from a coherent dipole source via electric-field-driven LC resonator metamaterials," Phys. Rev. Lett. **111**, 133901 (2013).

[36] P.H.Tichit, S. N.Burokur, and A.de Lustrac, "Ultradirective antenna via transformation optics," J. Appl. Phys. **105**, 104912 (2009).

[37] Y.Luo, J.Zhang, H.Chen, J.Huangfu, and L.Ran, "High directivity antenna with small antenna aperture," Appl. Phys. Lett. **95**, 193506 (2009).

[38] P.H.Tichit, S. N.Burokur, D.Germain, and A.de Lustrac, "Design and experimental demonstration of a high-directive emission with transformation optics," Phys. Rev. B **83**, 155108 (2011).

[39] Z.Wang, L.Shen, J.Chen, H.Wang, F.Yu, and H.Chen, "Highly directional small-size antenna designed with homogeneous transformation optics," Int. J.Antenn. Propag. **2014**, 91620 (2014).

[40] C. M.Segura, A.Dyke, H.Dyke, S.Haq, and Y.Hao, "Flat Luneburg lens via transformation optics for directive antenna applications," IEEE T. Antenn. Propag. **62**, 1945–53 (2014).

[41] Z.Xiao, J.Yao, and S.Yin, "High-directivity antenna array based on artifiicial electromagnetic metamaterials with low refractive index," Int. J. Antenn. Propag. **2015**, 294598 (2015).

[42] H. A.Madni, N.Aslam, S.Iqbal, S.Liu, and W. X.Jiang, "Design of homogeneous-material cloak and illusion devices for active and passive scatter with multi-folded transformation optics," J. Opt. Soc. Am. B **35**, 2399–404 (2018).

[43] H. A.Madni, B.Zheng, R.Zhu, L.Shen, H.Chen, Z.Xu, S.Dehdashti, Y.Zhao, and H.Wang, "Non-contact method to freely control the radiation patterns of antenna with multifolded transformation optics," Sci. Rep. **7**,13171(2017).

[44] Y.Lai, J.Ng, H. Y.Chen, D.Han, J.Xiao, Z.Q.Zhang, and C. T.Chan, "Illusion optics: the optical transformation of an object into another object," Phys. Rev. Lett. **102**, 253902 (2009).

[45] W. X.Jiang, H. F.Ma, Q.Cheng, and T. J.Cui, "Illusion media: generating virtual objects using realizable metamaterials," Appl. Phys. Lett. **96**, 121910 (2010).

[46] W. X. Jiang, H. F. Ma, Q. Cheng, and T. J. Cui, "Virtual conversion from metal object to dielectric object using metamaterials," Opt. Express. **18**, 11276–11281 (2010).

[47] W. X.Jiang, and T. J.Cui, "Radar illusion via metamaterials," Phys. Rev. E. **83,**026601 (2011).

[48] W. X.Jiang, T. J.Cui, X. M.Yang, H. F.Ma, and Q.Cheng, "Shrinking an arbitrary object as one desires using metamaterials," Appl. Phys. Lett. **98**, 204101 (2011).
[49] W. X.Jiang, C.W.Qiu, T.Han, S.Zhang, and T. J.Cui, "Creation of ghost illusions using wave dynamics in metamaterials," Adv. Funct. Mater. **23**, 4028 (2013).
[50] H. R.Shoorian, and M. S.Abrishamian, "Design of optical switches by illusion optics," J. Opt. **15**, 055107 (2013).
[51] D.Margusi, and H. R.Shoorian, "Remote nano-optical beam focusing lens by illusion optics," Appl. Phys. A **117**, 505–11(2014).
[52] H. A.Madni, B.Zheng, M.Akhtar, F.Jaleel, S.Liu, A.Mahboob, S.Iqbal, W. X.Jiang, and T. J.Cui, "A bi-functional illusion device based on transformation optics," J. Opt. **21**, 035104(2019).
[53] M.Liu, Z. L.Mei, X.Ma, and T. J.Cui, "Dc illusion and its experimental verification," Appl. Phys. Lett. **101**, 051905 (2012).
[54] X.He, and L.Wu, "Illusion thermodynamics: a camouflage technique changing an object into another one with arbitrary cross section," Appl. Phys. Lett. **105**,221904 (2014).
[55] F.Sun, S.Li, and S.He, "Translational illusion of acoustic sources by transformation acoustics," J. Acoust. Soc. Am. **142**, 1213–8 (2017).
[56] T.Yang, H.Chen, X.Luo, and H.Ma, "Superscatterer: enhancement of scattering with complementary media," Opt. Express. **16**, 18545–50 (2008).
[57] J.S.Mei, Q.Wu, K.Zhang, X.J.He, and Y.Wang, "Homogeneous illusion device exhibiting transformed and shifted scattering effect," Opt. Commun. **368**, 113–8 (2016).
[58] Y.Liu, Z.Liang, F.Liu, O.Diba, A.Lamb and J.Li, "Source illusion devices for flexural Lamb waves using elastic metasurfaces," Phys. Rev. Lett. **119**, 034301 (2017).
[59] S. Y. Wang and S. B. Liu, "Amplifying device created with isotropic dielectric layer," Chin. Phys. B. **23**, 024104 (2013).
[60] T.Li, M.Huang, J.Yang, W.Zhu, and J.Zeng, "A novel method for designing electromagnetic shrinking device with homogeneous material parameters," IEEE T. Magn. **49**, 5280–6 (2013).
[61] J.Ng, H.Chen, and C. T.Chan, "Metamaterial frequency selective super-absorber," Opt. Lett. **34**, 644–6 (2009).
[62] Y.Du, X. F.Zang, C.Shi, X. B.Ji, and Y. M.Zhu, "Shifting media induced super-resolution imaging," J. Opt. **17**, 025606 (2015).
[63] X. F.Zang, P. C.Huang, and Y. M.Zhu, "Optical illusions induced by rotating medium," Opt. Commun. **410**, 977–82 (2018).